\documentclass[aps,twocolumn,preprintnumbers,prd,superscriptaddress,nofootinbib,10pt]{revtex4-2}

\usepackage{amssymb,amsmath,amsthm,amsfonts}
\usepackage[usenames]{color}
\definecolor{darkred}{rgb}{0.5,0,0}

\usepackage{bm}
\usepackage{dcolumn}
\usepackage[utf8]{inputenc}
\usepackage{latexsym}
\usepackage{rotating}
\usepackage{longtable}

\usepackage{enumerate}
\usepackage{tensor,multirow}
\usepackage{url}
\usepackage[linktocpage]{hyperref}
\usepackage{float}
\usepackage{graphicx}
\usepackage{xcolor}

\begin{document}

\title{Quasinormal Modes in Modified Gravity \\ using Physics-Informed Neural Networks}

\author{Raimon Luna}
	\affiliation{Departamento de Astronom\'{i}a y Astrof\'{i}sica, Universitat de Val\`{e}ncia,
Dr. Moliner 50, 46100, Burjassot (Val\`{e}ncia), Spain}

\author{Daniela D. Doneva}
	\affiliation{Theoretical Astrophysics, Eberhard Karls University of T\"ubingen, T\"ubingen 72076, Germany}

\author{Jos\'e A. Font}
	\affiliation{Departamento de Astronom\'{i}a y Astrof\'{i}sica, Universitat de Val\`{e}ncia,
Dr. Moliner 50, 46100, Burjassot (Val\`{e}ncia), Spain}
	\affiliation{Observatori Astron\`{o}mic, Universitat de Val\`{e}ncia,
C/ Catedr\'{a}tico Jos\'{e} Beltr\'{a}n 2, 46980, Paterna (Val\`{e}ncia), Spain}

\author{Jr-Hua Lien}
	\affiliation{Theoretical Astrophysics, Eberhard Karls University of T\"ubingen, T\"ubingen 72076, Germany}

\author{Stoytcho S. Yazadjiev}
	\affiliation{Department of Theoretical Physics, Faculty of Physics, Sofia University, Sofia 1164, Bulgaria}
	\affiliation{Institute of Mathematics and Informatics, 	Bulgarian Academy of Sciences, 	Acad. G. Bonchev St. 8, Sofia 1113, Bulgaria}

\begin{abstract}
In this paper, we apply a novel approach based on physics-informed neural networks to the computation of quasinormal modes of black hole solutions in modified gravity. In particular, we focus on the case of Einstein-scalar-Gauss-Bonnet theory, with several choices of the coupling function between the scalar field and the Gauss-Bonnet invariant. This type of calculation introduces a number of challenges with respect to the case of General Relativity, mainly due to the extra complexity of the perturbation equations and to the fact that the background solution is known only numerically. The solution of these perturbation equations typically requires sophisticated numerical techniques that are not easy to develop in computational codes. We show that physics-informed neural networks have an accuracy which is comparable to traditional numerical methods in the case of numerical backgrounds, while being very simple to implement. Additionally, the use of GPU parallelization is straightforward thanks to the use of standard machine learning environments. 
\end{abstract}

\maketitle

%%%%%%%%%%%%%%%%%%%%%%%%%%%%%%%%%%%%%%%%%%%%%%%%%%%%%%%%%%%%%%%%%%%%%%%%%%%%%%
\section{Introduction}
\label{sec:Introduction}
%%%%%%%%%%%%%%%%%%%%%%%%%%%%%%%%%%%%%%%%%%%%%%%%%%%%%%%%%%%%%%%%%%%%%%%%%%%%%%

Black hole characteristic ringing has been long considered one of the ultimate tests of the laws of gravity at extreme curvature \cite{Kokkotas:1999bd,Berti:2009kk,Konoplya:2011qq,Berti:2018vdi}. Detecting at least two quasinormal mode (QNM) frequencies can already put severe constraints on any deviation from general relativity (GR) \cite{Dreyer:2003bv,Berti:2007zu} or the other way around -- it can provide an undoubted signature of a failure of Einstein's theory. Moreover, with the direct detection of gravitational waves by the Advanced LIGO and Advanced Virgo detectors~\cite{Aasi:2014jea,Acernese:2014hva} from compact binary coallescences \cite{LIGOScientific:2016aoc,LIGOScientific:2018mvr,LIGOScientific:2020ibl,KAGRA:2021vkt} and with the gradual upgrades of their sensitivities, observing high-precision post-merger black hole ringing becomes realistic \cite{Isi:2019aib,Correia:2023bfn,LIGOScientific:2020tif,LIGOScientific:2021sio,Carullo:2019flw,Cotesta:2022pci,Capano:2022zqm,Siegel:2023lxl}. This is especially true for the next generation of gravitational-wave detectors such as LISA, the Einstein Telescope~\cite{Punturo:2010zza}, and the Cosmic Explorer~\cite{2022ApJ...931...22S}. 

There are a number of challenges in properly interpreting the gravitational wave signal emitted during such characteristic ringing. This includes for example the fact that the post-merger signal is not a clear one and it will be contaminated with the nonlinear part of the actual merger of two black holes \cite{Clarke:2024lwi,Cheung:2023vki} igniting nonlinear modes in the spectrum \cite{Mitman:2022qdl,Cheung:2022rbm,Khera:2023oyf}. Overtones or higher harmonics can also play an important role \cite{Ota:2019bzl,Giesler:2019uxc,CalderonBustillo:2020rmh,Nee:2023osy}. Last but not least, one needs a precise theoretical calculation of the black hole QNM spectrum not only in GR but also in plausible  modifications of this theory. This is a difficult task on its own despite the fact that we are dealing with linear perturbation equations. In many cases of realistic beyond-Kerr black holes, simple analytical, or quasi-analytical approaches such as WKB \cite{Iyer:1986np, Iyer:1986nq}, are not accurate enough to compute the spectrum. Instead, the resulting differential equations governing the QNM frequencies are typically integrated numerically which is very involved since the equations are stiff \cite{Chandrasekhar:1975zza,Blazquez-Salcedo:2015ets}. Subtle adjustments of the numerical codes are required and the solution is typically done case by case \cite{Blazquez-Salcedo:2018pxo}. The picture worsens if we want to extend these results to the case of rapid rotation since in most of the realistic theories beyond GR separation of variables is not possible and one ends up with a system of partial differential equations instead.

To address the aforementioned obstacles, several well-designed approximations have been developed. These typically include an expansion in the coupling parameter for the beyond-GR theory, assuming that the GR corrections are small, and an expansion in terms of the rotational parameter \cite{Ayzenberg:2013wua,Cano:2021myl,Srivastava:2021imr,Wagle:2021tam,Pierini:2021jxd,Pierini:2022eim,Ghosh:2023etd,Wagle:2023fwl}. Sophisticated new methods for calculating the QNM frequencies using spectral decomposition are also being developed \cite{Jansen:2017oag,Chung:2023wkd,Blazquez-Salcedo:2023hwg}, showing promising potential but still posing a number of obstacles when dealing with beyond-GR theories. These approaches typically require important adjustments and the efforts behind each methodology/code development can be substantial. For example, the choice of a spectral basis can be quite critical for controlling the accuracy of spectral methods \cite{Jansen:2017oag,Chung:2023wkd,Blazquez-Salcedo:2023hwg}. Moreover, to avoid the contamination of the numerical solution with an ingoing wave component when performing direct numerical integration of the perturbations equations, advanced techniques such as complex scaling are typically applied that, however, require the adjustment of auxiliary parameters \cite{PhysRevA.44.3060,Samuelsson:2007hj,Blazquez-Salcedo:2012hdg}. In view of these difficulties it would be desirable to have a more general and straightforward  approach for QNM computation, even at the expense of some loss of accuracy, especially if it entailed less workforce. 

In this paper we explore the capabilities for QNM computation in beyond-GR theories of a relatively new such method based on machine learning and artificial neural networks (NNs). Neural networks have been used for the solution of ordinary and partial differential equations (PDEs) since the early 1990s (see \cite{Lee:1990, Meade:1994a, Meade:1994b, Yentis:1996, Lagaris:1997at, Lagaris:1997ap, Lagaris:2000, McFall:2009, Baymani:2010} and references therein). 
In particular, a class of NNs known as Physics-Informed Neural
Networks (PINNs) \cite{Raissi:2019, Nascimento:2020} have recently become a promising alternative to traditional numerical methods for solving PDEs. By capturing the behaviour of  physical systems described by their underlying PDEs, PINNs are now regarded as a competitive, straightforward approach for solving PDEs and have been applied to many different areas of mathematics and physics (e.g.~\cite{Jin:2020, Desai:2021, Mattheakis:2022, Ferrer:2024}). The working principle of PINNs relies on the ability of NNs to approximate any function by solving an optimization problem for the values of their weights and biases \cite{Hornik:1989}. As the functions involved in the definition of the NN are typically smooth, the gradients of the loss function can be efficiently computed by the chain rule by an algorithm known as backpropagation. Furthermore, in the particular case of PINNs, backpropagation is used to extract the derivatives of the interpolated function analytically.  

An additional advantage of PINNs is their ability to run on one or several graphics processing units (GPUs) without any significant modification of the code. As GPU computing is rapidly gaining importance, a large part of scientific computing hardware infrastructure is transitioning from the more traditional central processing units (CPUs) to clusters of GPUs. However, the adaptation of scientific numerical algorithms to the new devices typically needs a significant amount of work and time, even more so if parallelization is required. As PINNs can be implemented on standard machine learning environments such as \texttt{PyTorch} \cite{PyTorch}, they can be transferred to run on a GPU by changing only a few lines of code. All the adaptation and communication to the different type of hardware is handled in the background by the environment.  

PINNs have already been applied to the computation of black hole QNMs, both in the case of Schwarzschild \cite{Ovgun:2019yor, Ncube:2021jfu, Cornell:2022enn, Patel:2024wzo} and Kerr black holes \cite{Luna:2022rql, Cornell:2024azz}. It has been shown that its accuracy in calculating the quasinormal mode frequencies is good enough even for the next generation of gravitational wave detectors. So far, all previous applications of PINNs to QNMs have been performed on black hole solutions in GR, where the background metric is known analytically and the perturbation equations are relatively simple. In the present paper we build on our previous work \cite{Luna:2022rql} and go a step further by developing a PINN that calculates for the first time the QNMs of black holes beyond GR. The obstacles we had to overcome are related to the significantly more complicated perturbation equations compared to GR and also to the fact that the background black hole solution that is perturbed is known only numerically. 

The modified gravity theory under consideration will be the Einstein-scalar-Gauss-Bonnet (EsGB) gravity. This choice is motivated by the fact that black hole perturbations and QNMs have been extensively studied in the literature for this theory -- both in the static limit when solving the complete perturbation equations \cite{Blazquez-Salcedo:2015ets,Blazquez-Salcedo:2016enn,Blazquez-Salcedo:2020rhf,Blazquez-Salcedo:2022omw,Blazquez-Salcedo:2023hwg} or in the slow rotating approximation \cite{Pierini:2021jxd,Pierini:2022eim}. Particularly interesting for our studies is the case of axial perturbations of scalarized black holes considered in \cite{Blazquez-Salcedo:2020rhf} where the perturbation equation was derived in a very concise form. This allows us to borrow the methodology from previous QNM studies using PINNs \cite{Luna:2022rql} to better estimate the potential of using neural networks for calculating QNMs beyond GR. 

The paper is structured as follows: In Section \ref{sec:Setup} we introduce the perturbation equations for EsGB gravity and the procedure that we use to adapt them to the PINN. We describe the PINN itself and the way it operates on the numerical background data in Section \ref{sec:PINN}. We also perform in this section a series of tests to properly measure the accuracy of the method. Finally, in Section \ref{sec:Results} we present the quasinormal mode results for families of black hole solutions constructed on a variety of models. Our conclusions are outlined in Section \ref{sec:Discussion}.

%%%%%%%%%%%%%%%%%%%%%%%%%%%%%%%%%%%%%%%%%%%%%%%%%%%%%%%%%%%%%%%%%%%%%%%%%%%%%%
\section{Setup}
\label{sec:Setup}
%%%%%%%%%%%%%%%%%%%%%%%%%%%%%%%%%%%%%%%%%%%%%%%%%%%%%%%%%%%%%%%%%%%%%%%%%%%%%%
%
\subsection{Gauss-Bonnet Gravity}

This paper considers solutions of EsGB theories, which are described by the action
\begin{equation}
    S=\frac{1}{16\pi}\int d^4x \sqrt{-g} \left[R - 2\nabla_\mu \varphi \nabla^\mu \varphi + \lambda^2 f(\varphi) {\cal R}^2_{GB} \right] \, ,
    \label{eqn:GB_action}
\end{equation}
where the scalar field $\varphi$ couples via the coupling function $f(\varphi)$ to the Gauss-Bonnet invariant ${\cal R}^2_{GB}$, which is defined as 
\begin{equation}
    {\cal R}^2_{GB} = R^2 - 4 R_{\mu\nu} R^{\mu\nu} + R_{\mu\nu\alpha\beta} R^{\mu\nu\alpha\beta}\, .
\end{equation}
Variation of Eq.~(\ref{eqn:GB_action}) with respect to the metric $g_{\mu\nu}$ and the scalar $\varphi$, respectively, leads to the equations of motion
\begin{equation}
\begin{split}
    R_{\mu\nu} - \frac{1}{2}R g_{\mu\nu} + \Gamma_{\mu\nu} &= 2\nabla_\mu \varphi \nabla_\nu \varphi -  g_{\mu\nu} \nabla_\alpha \varphi \nabla^\alpha \varphi \, ,\\ \nabla_\alpha \nabla^\alpha \varphi &= - \frac{\lambda^2}{4} \frac{df(\varphi)}{d\varphi} {\cal R}^2_{GB} \, .
    \label{eqn:eom}
\end{split}
\end{equation}
with the tensors $\Gamma_{\mu\nu}$ and $\Psi_\mu$ being defined respectively as
\begin{equation}
\begin{split}
    \Gamma_{\mu\nu} &= - R(\nabla_\mu \Psi_\nu + \nabla_\nu \Psi_\mu ) - 4 g_{\mu\nu} R^{\alpha\beta} \nabla_\alpha \Psi_\beta \\ & + 4R_{\mu\alpha} \nabla^\alpha \Psi_\nu + 4R_{\nu\alpha} \nabla^\alpha \Psi_\mu + 4 R^{\beta}_{\;\mu\alpha\nu} \nabla^\alpha \Psi_\beta \\ & - 4 \nabla^\alpha \Psi_\alpha \left(R_{\mu\nu} - \frac{1}{2}R g_{\mu\nu}\right) \, ,\\
    \Psi_\mu &= \lambda^2 \frac{df(\varphi)}{d\varphi} \nabla_\mu \varphi \, .
\end{split}
\end{equation}
\subsection{Scalar field coupling}

We will consider two coupling functions $f(\varphi)$, namely
\begin{eqnarray}
    &&f_2(\varphi) = \frac{1}{2\beta} \left(1 - e^{-\beta \varphi^2}\right)\,, \label{eq:f_2}\\
    &&f_4(\varphi) = \frac{1}{4\beta} \left(1 - e^{-\beta \varphi^4}\right) \, . \label{eq:f_4}
\end{eqnarray}
Even though they look similar, $f_2(\varphi)$ and $f_4(\varphi)$ have very distinct phenomenology. First of all let us note that the exponential form in 
Eqs.~\eqref{eq:f_2} and \eqref{eq:f_4} is taken for numerical convenience. What matters most for the characteristic features of the spectrum of solutions is the leading-order expansion of the coupling for small scalar fields. Thus, in the limit $\varphi \rightarrow 0$ we have $f_2(\varphi) \sim \varphi^2$ and $f_4(\varphi) \sim \varphi^4$. Following this, it is clear that for both couplings
\begin{equation}
    \left .\frac{\partial f_{2,4}}{\partial \varphi} \right|_{\varphi = 0} = 0\,.
\end{equation}
Therefore, based on the field equations \eqref{eqn:eom} one can conclude that GR (with $\varphi = 0$) is always a solution. In the case of $f_2(\varphi)$ one also has that
\begin{equation}
\left. \frac{\partial^2 f_2}{\partial \varphi^2} \right|_{\varphi = 0} > 0 \,,
\end{equation}
that eventually renders the Schwarzschild black hole unstable for small enough masses (or equivalently large dimensional coupling $\lambda$) and gives rise to linearly stable scalarized black hole solutions \cite{Doneva:2022ewd}.

For the second coupling the picture changes since
\begin{equation}
   \left. \frac{\partial^2 f_4}{\partial \varphi^2} \right|_{\varphi = 0} = 0\,.
\end{equation}
In that case, the Schwarzschild black hole is always linearly stable but in addition, linearly stable scalarized black holes can exist as well being thermodynamically preferred over GR for a wide range of the parameters \cite{Doneva:2021tvn}. A transition between the two can happen through a large nonlinear perturbation. Thus, we call this case nonlinear scalarization.

In the QNM calculation reported in this work we will employ both couplings, with $f_2(\varphi)$ being our test case for estimating the error of the PINN method based on previous results \cite{Blazquez-Salcedo:2020rhf}, while the QNM calculation for $f_4$ are original results calculated for the first time in the present paper.

\subsection{Black hole static solutions}

The background static black hole solutions of (\ref{eqn:eom}) are spherically symmetric assuming the following ansatz for the metric
\begin{equation} \label{eqn:BG_metric}
    ds^2= - e^{2\mu_0(r)}dt^2 + e^{2\nu_0(r)} dr^2 + r^2 (d\theta^2 + \sin^2\theta d\phi^2 ) \, .
\end{equation}
Therefore, the solutions will be entirely described by the radial profiles of $\mu_0(r)$, $\nu_0(r)$ and $\varphi_0(r)$. The subscript $(\cdot)_0$ from now on will denote the background unperturbed quantities. 

When the metric ansatz \eqref{eqn:BG_metric} is plugged in the field equations \eqref{eqn:eom}, one can easily derive the dimensionally reduced field equations. The boundary conditions are the natural ones -- regularity at the black hole horizon and asymptotic flatness at infinity. Since the background solutions are not the focus on the present paper we refer the reader to \cite{Doneva:2017bvd} for further details. The black holes are calculated numerically using the code developed in \cite{Doneva:2017bvd,Doneva:2021tvn}. The so-obtained solutions are then transferred to the PINN code to compute the appropriate coefficients in the linear perturbation equations.

\subsection{Axial Perturbation Equations}

Axial perturbations of spherical EsGB black holes \cite{Blazquez-Salcedo:2020rhf} are described by the (linear) equation
\begin{equation} 
\begin{split}
    &\frac{\partial^2 \Psi}{\partial \tilde{r}_*^2} + \left[\frac{1}{2 \mathcal{S}_0} \, \frac{\partial^2 \mathcal{S}_0}{\partial \tilde{r}_*^2} - \frac{3}{4 \mathcal{S}_0^2}\, \left(\frac{\partial \mathcal{S}_0}{\partial \tilde{r}_*}\right)^2 - \frac{e^{\sigma_0}}{r \mathcal{S}_0} \, \frac{\partial \mathcal{S}_0}{\partial \tilde{r}_*} \right. \\ &\left. + \left(r \frac{\partial \sigma_0}{\partial r} - 2\right) \, \frac{e^{2 \sigma_0}}{r^2} - (\ell-1)(\ell+2) \, \frac{e^{2 \mu_0} \mathcal{S}_0}{r^2 \mathcal{W}_0}\right] \Psi \\ &= \frac{\mathcal{S}_0}{\mathcal{P}_0} \, \frac{\partial^2 \Psi}{ {\partial t}^2} \, ,
    \label{eqn:pert_tortoise}
\end{split}
\end{equation} 
where the background auxiliary functions $\sigma_0$, $\mathcal{P}_0$, $\mathcal{W}_0$ and $\mathcal{S}_0$ are defined as
\begin{equation} 
\begin{split}
\sigma_0(r) &= \mu_0 - \nu_0 \, , \\
\mathcal{P}_0(r) &= 1 - 4\lambda^2 \, \frac{\partial f}{\partial r} \, \frac{\partial \mu_0}{\partial r} \, e^{-2\nu_0} \, ,  \\
\mathcal{W}_0(r) &= 1 - 4\lambda^2 \, \frac{\partial f}{\partial r} \, \frac{e^{-2\nu_0}}{r} \,  , \\
\mathcal{S}_0(r) &= 1 - 4\lambda^2 \left( \frac{\partial^2 f}{\partial r^2} - \, \frac{\partial f}{\partial r} \, \frac{\partial \nu_0}{\partial r} \, \right) e^{-2\nu_0} \, .
\label{eqn:PWS}
\end{split}
\end{equation} 
We should note that the perturbation equation \eqref{eqn:pert_tortoise} is a wave-like equation for $\Psi$ only when $\mathcal{S}_0/\mathcal{P}_0>0$. In case this ratio turns negative, Eq. \eqref{eqn:pert_tortoise} changes character from hyperbolic to elliptic. In this regime, the quasinormal modes are not well defined and we will exclude such background black hole solutions from our analysis below.  

In the form of (\ref{eqn:pert_tortoise}), the perturbation equation involves two radial coordinates simultaneously, namely the Schwarzschild-like coordinate $r$ and the Regge-Wheeler coordinate $\tilde r_*$. Both coordinates are related to each other by
\begin{equation} 
    \frac{\partial}{\partial {\tilde{r}_*} } = e^{\sigma_0} \frac{\partial}{ \partial r} \; .
\end{equation} 
Rewriting all derivatives in terms of $r$, as well as taking a Fourier mode in the form $\Psi(r, t) = e^{-i \omega t} \psi(r)$, leads to the equation
\begin{equation} 
\begin{split}
    & \frac{\partial^2 \psi}{\partial r^2} +  \frac{\partial \sigma_0}{\partial r} \, \frac{\partial \psi}{\partial r} + \frac{1}{2  \mathcal{S}_0} \, \left[ \frac{\partial^2  \mathcal{S}_0}{\partial r^2}  -  \frac{3}{2 \mathcal{S}_0} \, \left( \frac{\partial \mathcal{S}_0}{\partial r} \right)^2 \right. \\ & + \left( \frac{\partial \sigma_0}{\partial r}  - \frac2{r} \right) \, \frac{\partial \mathcal{S}_0}{\partial r} + \left(r \frac{\partial \sigma_0}{\partial r} - 2 \right) \, \frac{2\mathcal{S}_0}{r^2}  \\ &\left. - (\ell-1)(\ell+2) \, \frac{2 \, \mathcal{S}_0^2 \, e^{2\nu_0}}{r^2 \mathcal{W}_0 } + \frac{2 \omega^2 \, \mathcal{S}_0^2 \, e^{-2 \sigma_0}}{\mathcal{P}_0} \right] \psi = 0 \; .
    \label{eqn:pert_r}
\end{split} 
\end{equation} 
The coordinate $r$ extends from the singularity at $r = 0$ all the way to space-like infinity $r \to \infty$, with an event horizon at $r = r_H$. The perturbation equations have to be solved on the black hole exterior, i.e., in the domain $r \in [r_H, \infty)$. Working with computational domains of infinite extension is usually impractical for numerical techniques, so it is convenient to compactify the radial coordinate as
\begin{equation}
    x = \frac{r_H}{r}\; ,
\end{equation}
which reduces the domain to $x \in [0, 1]$. In terms of the compactified radial coordinate $x$,  Eq.~(\ref{eqn:pert_r}) becomes
\begin{equation}
\begin{split}
    & x^4 \psi ''+\left(2 x^3+x^4 \sigma_0' \right) \psi ' + \frac{1}{2 \mathcal{S}_0} \left[ x^4 \mathcal{S}_0''-\frac{3 x^4}{2 \mathcal{S}_0}  (\mathcal{S}_0')^2 \right . \\ & +\left(4 x^3+x^4 \sigma_0' \right) \mathcal{S}_0'  -2 x^2 \left(2+x \sigma_0' \right) \mathcal{S}_0 \\ & \left. - (\ell-1) (\ell+2) \frac{2 x^2 \, \mathcal{S}_0^2 \, e^{2 \nu_0 }}{\mathcal{W}_0} + \frac{2 \Omega^2 \, \mathcal{S}_0^2 \, e^{-2 \sigma_0 }}{\mathcal{P}_0}  \right]\psi = 0 \, ,
    \label{eqn:pert_x}
\end{split}
\end{equation}
where the primes denote derivatives with respect to $x$, and we have introduced the scale-invariant parameter
\begin{equation}
\Omega = r_H \, \omega    
\end{equation}
as the frequency normalized by the horizon radius, which will be very useful for the numerical treatment of the equations. Additionally, all functions that have to be treated numerically should not diverge and be as smooth as possible inside the computational domain. This is particularly important in the computation of QNM, as we are searching for regular solutions for the perturbation. For this reason, it is useful to consider the equations in the GR limit to analyse the analytic structure of the functions. Indeed, in the Schwarzschild case the functions $\mu_0(x)$ and $\nu_0(x)$ take the simple form $\mu_\text{Schw}(x) = - \nu_\text{Schw}(x) = \frac12 \log(1-x)$, which have a logarithmic singularity at $x=1$. For this reason, it is fundamental to redefine 
\begin{equation}
    \mathcal{M}_0(x) \equiv e^{2\mu_0(x)}, \quad \mathcal{N}_0(x) \equiv e^{-2\nu_0(x)} \; ,
\end{equation}
so we can work in terms of the regular functions $\mathcal{M}_\text{Schw}(x) = \mathcal{N}_\text{Schw}(x) = 1-x$. Imposing boundary conditions for a wave that is ingoing at the horizon and outgoing at spatial infinity, we get that $\psi$ can be expressed in the form
\begin{equation}
    \psi(x) = \left(\frac{1}{x}-1\right)^{-i \gamma  \Omega} \left(\frac{1}{x}\right)^{i (\gamma +\delta ) \Omega} e^{i \left(\frac{1}{x}-1\right) \Omega} g(x)\; ,
\end{equation}
where $g(x)$ is a regular function for $x \in [0,1]$ and we defined the exponents $\gamma$ and $\delta$ as
\begin{equation}
\begin{split}
    \gamma &= - \lim_{x \to 1} \left[ \frac{\mathcal{S}_0}{\mathcal{P}_0} \frac{1}{x^2 \; \partial_x \sqrt{\mathcal{M}_0\mathcal{N}_0}} \right] = \frac{\mathcal{S}_0(r_H)}{2 \, \kappa \, r_H \, \mathcal{P}_0(r_H)}, \\\quad \delta &= \lim_{x \to 0} \frac{1 - \sqrt{\mathcal{M}_0\mathcal{N}_0}}{x} = \frac{2M}{r_H}.
\end{split}
\end{equation}
In the previous equations $\kappa$ and $M$ are the surface gravity and the mass of the black hole solution, respectively. Now, after the change of variable, the differential equation for $\psi(x)$ becomes a second order differential equation for $g(x)$ whose coefficients depend on the combination $i \, \Omega$, as
\begin{equation}
    \frac{1}{4x^2(x-1)^3}\sum_{n = 0}^2 \; \sum_{m = 0}^{2-n} \; A_{n,m} \; g^{(n)}(x) \, (i \, \Omega)^m = 0\; .
    \label{eqn:PINN_eqn}
\end{equation}
Here $g^{(n)}(x)$ is the $n$-th derivative of the function $g(x)$. The real coefficients $A_{n,m}(x)$ are 
\begin{equation}
\begin{split}
    A_{0,0} =&   - x^2 (x - 1)^2 \; [4(\ell-1)(\ell+2) \mathcal{M}_0 \, \mathcal{S}_0^3 \\ &- x(8 \mathcal{N}_0 \, \mathcal{S}_0' + x \mathcal{N}_0' \, \mathcal{S}_0' + 2 x \mathcal{N}_0 \, \mathcal{S}_0'') \, \mathcal{M}_0 \, \mathcal{W}_0 \, \mathcal{S}_0 \\ &+ x (2 \mathcal{S}_0 - x \mathcal{S}_0') \, \mathcal{M}_0' \, \mathcal{N}_0 \, \mathcal{W}_0 \, \mathcal{S}_0 \\ & + 3 x^2 \mathcal{M}_0 \, \mathcal{N}_0 \, \mathcal{W}_0 \, \mathcal{S}_0'^2  \\ & + 2(4 \mathcal{N}_0 + x \mathcal{N}_0') \, \mathcal{M}_0 \, \mathcal{W}_0 \, \mathcal{S}_0^2 ] \, \mathcal{P}_0 \\
    A_{0,1} =&  -2 x^2 \, [2 (\delta + (x^2 - 2 x)(\gamma + \delta)) \mathcal{M}_0 \, \mathcal{N}_0  \\ &+ (x-1) B(x) (\mathcal{M}_0 \, \mathcal{N}_0)'] \, \mathcal{P}_0 \, \mathcal{W}_0 \, \mathcal{S}_0^2 \, , \\
    A_{0,2} =& + 4 \, [ B^2(x) \, \mathcal{M}_0 \, \mathcal{N}_0 \, \mathcal{P}_0 - (x-1)^2 \, \mathcal{S}_0  ] \, \mathcal{W}_0 \, \mathcal{S}_0^2 \, , \\
    A_{1,0} =&  +2 x^3 (x-1)^2 \, \left[x (\mathcal{M}_0 \, \mathcal{N}_0)' + 4 \mathcal{M}_0 \mathcal{N}_0\right] \, \mathcal{P}_0 \, \mathcal{W}_0 \, \mathcal{S}_0^2 \, , \\
    A_{1,1} =&  -8 x^2 (x-1) \, B(x) \, \mathcal{M}_0 \, \mathcal{N}_0 \, \mathcal{P}_0 \, \mathcal{W}_0 \, \mathcal{S}_0^2 \, , \\
    A_{2,0} =& + 4 x^4 (x-1)^2  \, \mathcal{M}_0 \, \mathcal{N}_0 \, \mathcal{P}_0 \, \mathcal{W}_0 \, \mathcal{S}_0^2 \, ,
\label{eqn:A_coefs}
\end{split}
\end{equation}
where we have defined the auxiliary function $B(x)$ as
\begin{equation}
    B(x) = x^2 (\gamma +\delta) + x (1 - \delta) - 1 \, . 
\end{equation}
The normalization factor of $4x^2(x-1)^3$ in (\ref{eqn:PINN_eqn}) is motivated by the Schwarzschild limit, where it appears as a common multiplying factor in the $A_{n, m}$. Cancelling this factor out has been shown to make numerical approaches more stable \cite{Jansen:2017oag}. It turns out that the necessity of dividing the entire equation by this normalizing factor remains even in EsGB solutions beyond Schwarzschild. 

The background functions $\mathcal{P}_0$, $\mathcal{W}_0$ and $\mathcal{S}_0$ now become, in terms of the compactified variable $x$,
\begin{equation}
\begin{split}
    \mathcal{P}_0 =& 1 - 2 x^4 \left(\frac{\lambda}{r_H} \right)^2 \frac{\mathcal{M}'_0 \, \mathcal{N}_0 }{ \mathcal{M}_0}  \, f' \, ,\\ 
    \mathcal{W}_0 =& 1 + 4 x^3 \left(\frac{\lambda}{r_H} \right)^2 \mathcal{N}_0 \, f'\, ,\\
    \mathcal{S}_0 =& 1 - 2 x^3 \left(\frac{\lambda}{r_H} \right)^2 \left[x \mathcal{N}'_0 f' + 2 \mathcal{N}_0 \left( 2f' + x f'' \right) \right] \, .\\
\end{split}
\end{equation}
Writing the perturbation equations in this form is especially convenient as the coefficients $A_{n,m}$ can be computed numerically beforehand on any computational grid, with any differentiation technique, without the need of any machine learning training. This reduces drastically the time and computational cost of the method.

\subsection{The Schwarzschild case}

In the particular case of Schwarzschild, we have the coupling $\lambda_\text{Schw} = 0$, the background functions $\mathcal{S}_\text{Schw} = \mathcal{W}_\text{Schw} = \mathcal{P}_\text{Schw} = 1$ and the metric functions $\mu_\text{Schw}(r) = -\nu_\text{Schw}(r) = \frac{1}2 \log \left(1- \frac{r_H}{r}\right)$. It is easy to check that in this particular case Eq.~(\ref{eqn:pert_tortoise}) becomes the well-known Schrödinger-type equation
\begin{equation} 
\begin{split}
    &\left( \frac{\partial^2}{\partial \tilde{r}_*^2} - \frac{\partial^2}{\partial t^2} + V_l(r) \right) \Psi = 0 \, , \\ &V_l (r) = \left( 1- \frac{r_H}{r} \right) \left( - \frac{\ell(\ell+1)}{r^2} + \frac{3r_H}{r^3} \right) \, .
\end{split}
\end{equation}
On the other hand, we can check that from Eq.~(\ref{eqn:pert_r}) we recover the Regge-Wheeler equation
\begin{equation} 
\begin{split}
    & r(r - r_H) \frac{\partial^2 \psi}{\partial r^2} + r_H \frac{\partial \psi}{\partial r} \\ + &\left[ \frac{\omega^2 r^3}{r- r_H} - \ell(\ell+1) + \frac{3r_H}{r}\right] \psi = 0\; .
\end{split} 
\end{equation} 
Finally, Eq.~(\ref{eqn:PINN_eqn}), evaluated for the Schwarzschild solution values of the background functions $\mathcal{M}_\text{Schw} = \mathcal{N}_\text{Schw} = 1 - x$ becomes the compactified form
\begin{equation}
\begin{split}
    &[\ell(\ell + 1) -3 x-4 i x \Omega -4 (1+x) \Omega ^2] \, g(x) \\ + &[3x^2 - 2x + (2 -4 x^2) i \Omega ] \, g'(x) \\ + &(x - 1) \, x^2 \, g''(x) = 0\,,
\end{split}
\end{equation}
often used when solving for QNM numerically, for instance with spectral decomposition methods \cite{Jansen:2017oag}.

%%%%%%%%%%%%%%%%%%%%%%%%%%%%%%%%%%%%%%%%%%%%%%%%%%%%%%%%%%%%%%%%%%%%%%%%%%%%%%
\section{The PINN}
\label{sec:PINN}
%%%%%%%%%%%%%%%%%%%%%%%%%%%%%%%%%%%%%%%%%%%%%%%%%%%%%%%%%%%%%%%%%%%%%%%%%%%%%%

\subsection{Numerical Approach}

The computation of the numerical values of the QNM has two main parts:
\begin{itemize}
    \item Extraction of background functions $A_{n,m}(x)$
    \item Training of the Physics-Informed Neural Network
\end{itemize}

\begin{figure}[thpb]
\begin{center}
\includegraphics[width=0.49\textwidth]{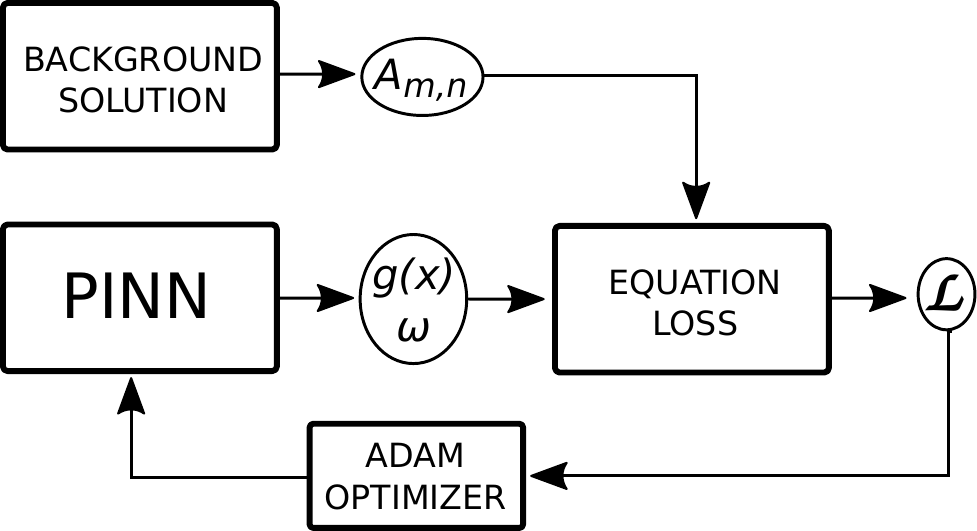}
\caption{Scheme of the PINN setup, showing both the backgound data reduction and the machine learning training to extract the QNM frequencies.
\label{fig:Setup_Scheme}}
\end{center}
\end{figure}

The first part comprises the process of reduction of the numerical background solutions and the computation of the coefficients $A_{n,m}$ from (\ref{eqn:A_coefs}). This part does not involve any machine learning technique, and can be performed without keeping track of the computation graphs or gradients. As mentioned before this speeds up the process drastically. In practice, the grid functions $\mu_0(r)$, $\nu_0(r)$ and $\varphi_0(r)$ are interpolated by cubic splines in order to evaluate the coefficients $A_{n,m}$ on a set of PINN collocation points $x_i \in [0, 1] \, \forall i$. Splines also allow for the fast extraction of the derivatives that are needed in the expressions for $A_{n,m}$. These coefficients contain all the physical information on the background solution that will then be used by the machine learning model. In practice, we choose $x_i$ to be a set of 1000 uniformly distributed points between $x = \epsilon$ and $x = 1 - \epsilon$. The regulator $\epsilon = 10^{-2}$ avoids numerical indeterminations due to the vanishing normalization factor in Eq.~(\ref{eqn:PINN_eqn}).

The second part contains all the machine learning techniques inside the \texttt{PyTorch} \cite{PyTorch} environment, and involves the actual physics-informed neural network which acts as a universal approximant of $x \to g(x)$. This network contains as trainable parameters its own weights and biases, as well as the complex value for the normalized frequency $\Omega$. The network has an input layer with a single neuron, accepting $x$, and an output layer with two neurons, approximating the real and imaginary part of $g(x)$. All numerical values inside the network are real, as the activation functions do not behave well with complex values. We use 3 hidden layers of 200 neurons each, with GELU activation functions between them. Moreover, we use a hard-enforced normalization condition to impose $g(1) = 1$, by the relation
\begin{equation}
    g(x) = \left(e^{x-1} - 1\right) \mathcal{G}(x) + 1\, ,
\end{equation}
where $\mathcal{G}(x)$ is the complex output of the network. The process is explained in more detail in \cite{Luna:2022rql}.

The PINN-approximated function $g(x)$, together with its derivatives obtained by automatic differentiation, are then combined with the $A_{n,m}(x)$ coefficients in the perturbation equation. The loss function $\mathcal{L}$ is the complex magnitude of the right-hand side of (\ref{eqn:PINN_eqn}), averaged over the set of PINN collocation points. This loss is then minimized by an Adam optimizer \cite{Adam} with a learning rate of $r_l = 10^{-3}$ over 1000 epochs, which updates the weights, biases and $\Omega$ of the PINN. With this process, the approximations of $g(x)$ and $\Omega$ converge to the values of the dominant QNM of the system. The process is graphically represented in FIG. \ref{fig:Setup_Scheme}. 

The weights of the PINN are initialized on normally distributed values with zero mean and standard deviation of $\sigma = 0.1$, while biases are initialized to zero. The (normalized) frequency is initialized at $\Omega_0 = 0.7 -0.1i$, as this is close to the fundamental Schwarzschild QNM for $\ell=2$. When computing QNM on a one-parameter family of black hole solutions of the same theory (usually parametrized by their horizon radius $r_H$), we initialize the network as explained above and start training the network on the largest $r_H$, which is typically simpler. We then proceed by computing the QNM of the remaining solutions by decreasing order in $r_H$. In each consecutive mode we take the results of the previous training as starting values for the trainable parameters, without initializing the network again. Several one-parameter families of solutions can be 
trained simultaneously on the same GPU. All combined QNM results in this paper (1873 quasinormal modes in total) were created in approximately 2 hours in a single Nvidia RTX 4090 24GB GPU.

\subsection{Testing the PINN}
\label{subsec:testingPINN}

The accuracy and reliability of the PINN are measured in accordance with its results on previously known values of the fundamental axial QNM for $\ell=2$, both in GR and in EsGB theories\footnote{{Note that unlike in GR, in EsGB gravity the axial and polar QNMs have distinct frequencies \cite{Blazquez-Salcedo:2020caw}.}}. We also relate the value of the loss function at the end of the training process to the mode accuracy, which provides a very straightforward way to gauge the reliability of the numerical values obtained on previously unknown QNM. 

\begin{widetext}

\begin{table}[thpb]
\begin{center}
\setlength\tabcolsep{8pt}
  \begin{tabular}{|c||c|c|c|c|}
    \hline
    \multicolumn{5}{|c|}{Real part of $\omega$} \\ \hline \hline
        $r_H$ & Original & PINN & \% error & loss \\ \hline \hline 
        1.174 & 0.63661 & 0.63662 & 0.0013 & 0.0050 \\ \hline
        1.10 & 0.65956 & 0.65957 & 0.0016 & 0.0048 \\ \hline
        1.00 & 0.69767 & 0.69772 & 0.0083 & 0.0053 \\ \hline
        0.90 & 0.74966 & 0.74968 & 0.0027 & 0.0046 \\ \hline
        0.80 & 0.82837 & 0.82827 & 0.0123 & 0.0069 \\ \hline
        0.70 & 0.94138 & 0.94159 & 0.0224 & 0.0082 \\ \hline
        0.60 & 1.09520 & 1.09547 & 0.0244 & 0.0129 \\ \hline
        0.50 & 1.30678 & 1.30669 & 0.0069 & 0.0157 \\ \hline
        0.40 & 1.60881 & 1.60854 & 0.0170 & 0.0152 \\ \hline
        0.30 & 2.06030 & 2.06493 & 0.2247 & 0.0791 \\ \hline
        0.25 & 2.37548 & 2.49225 & 4.9157 & 0.2441 \\ \hline
        0.20 & 2.78651 & 4.25151 & 52.5745 & 0.7993 \\ \hline
    \multicolumn{5}{c}{} 
  \end{tabular}
  $\qquad$
  \begin{tabular}{|c||c|c|c|c|}
    \hline
    \multicolumn{5}{|c|}{Imaginary part of $\omega$} \\ \hline \hline
        $r_H$ & Original & PINN & \% error & loss \\ \hline \hline   
        1.174 & -0.15156 & -0.15165 & 0.0588 & 0.0050 \\ \hline
        1.10 & -0.15946 & -0.15946 & 0.0018 & 0.0048 \\ \hline
        1.00 & -0.16963 & -0.16914 & 0.2861 & 0.0053 \\ \hline
        0.90 & -0.17341 & -0.17314 & 0.1581 & 0.0046 \\ \hline
        0.80 & -0.17252 & -0.17245 & 0.0420 & 0.0069 \\ \hline
        0.70 & -0.17865 & -0.17874 & 0.0511 & 0.0082 \\ \hline
        0.60 & -0.19721 & -0.19733 & 0.0583 & 0.0129 \\ \hline
        0.50 & -0.23101 & -0.23080 & 0.0905 & 0.0157 \\ \hline
        0.40 & -0.28355 & -0.28367 & 0.0425 & 0.0152 \\ \hline
        0.30 & -0.34678 & -0.34863 & 0.5323 & 0.0791 \\ \hline
        0.25 & -0.35989 & -0.49251 & 36.8501 & 0.2441 \\ \hline
        0.20 & -0.32339 & -1.33521 & 312.8840 & 0.7993 \\ \hline
    \multicolumn{5}{c}{} 
  \end{tabular}
  \caption{Comparison between the modes in \cite{Blazquez-Salcedo:2020rhf} and those computed by the PINN for scalarized black holes with coupling function $f_2(\varphi_0)$   and $\beta=6$.
  \label{tab:real_imag}}
\end{center}
\end{table}

\end{widetext}

\begin{figure}[thpb]
\begin{center}
\includegraphics[width=0.46\textwidth]{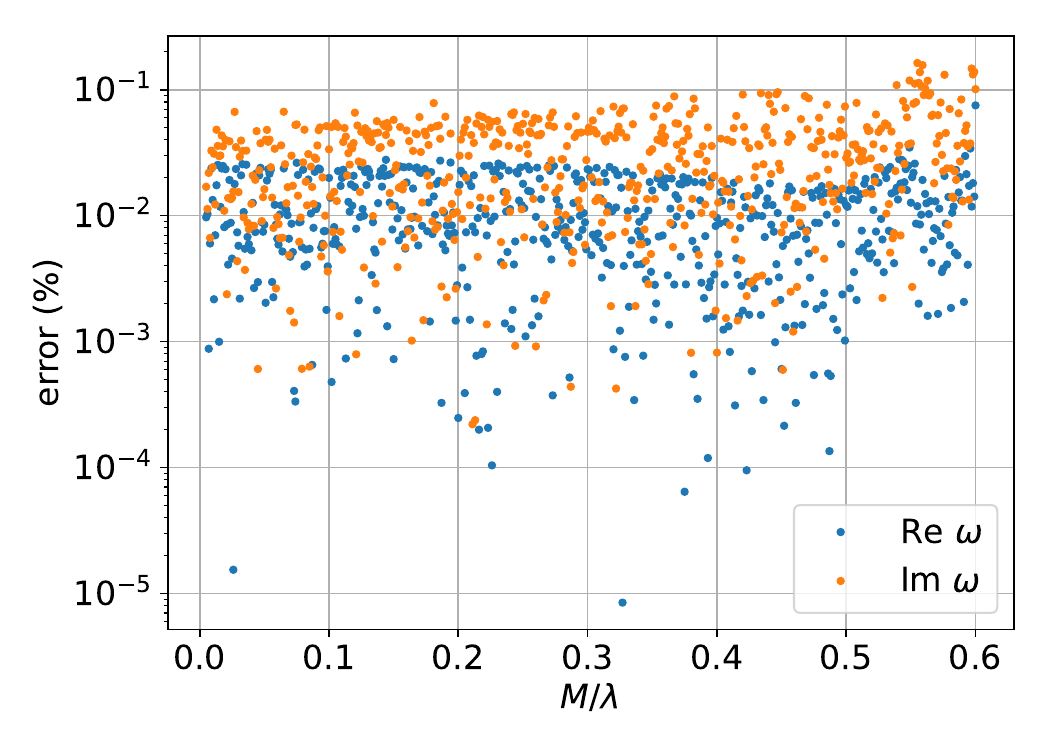}
\caption{Neural network error for the GR modes, as a function of the mass $M = r_H / 2$.
\label{fig:rh_vs_error}}
\end{center}
\end{figure}

The simplest test, both for the PINN and the numerical accuracy of the background solution, is to extract modes of the Schwarzschild solution with different horizon radii $r_H$. In this case the fundamental frequency for $\ell=2$ is very well-known to be $\Omega_\textbf{Schw} = 0.747343 - 0.177925 \, i$, so it is easy to compare. Notice that, in the particular case of the Schwarzschild solution, $r_H = 2 M$.  In FIG. \ref{fig:rh_vs_error} we show the percentual accuracy of the PINN on GR modes, for 596 values of $r_H$ ranging from 0.01 to 1.2.

Next, we compare the QNM extracted with the PINN to a selection of values from \cite{Blazquez-Salcedo:2020rhf} calculated for scalarized black holes having the coupling $f_2(\varphi_0)$ (see Eq.~\eqref{eq:f_2}) with $\beta=6$. The results are summarized in Table \ref{tab:real_imag}. All the modes agree well below the percentual level, with the only exception of the solutions with $r_H \leq 0.25$, where the accuracy falls dramatically. In this case, the error could also be caused by the proximity to the point of hyperbolicity loss in the equations of motion (\ref{eqn:eom}), as well as the possible decrease in the effective resolution of the background.

\begin{figure}[thpb]
\begin{center}
\includegraphics[width=0.49\textwidth]{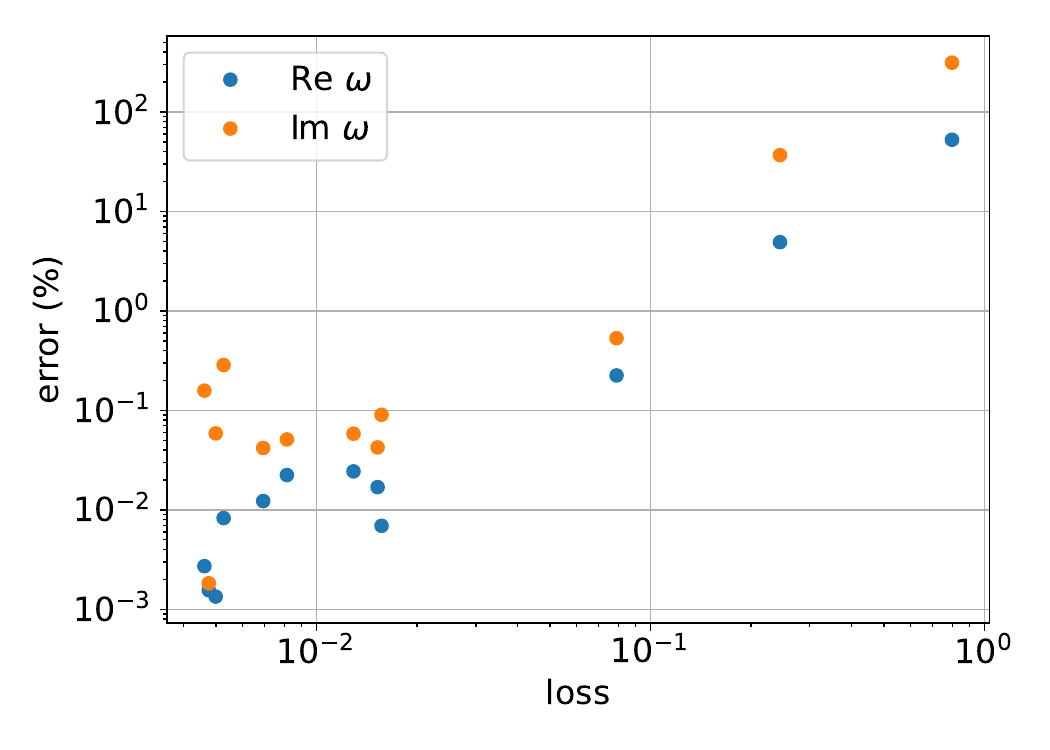}
\caption{{Neural network loss vs accuracy of the reference modes in \cite{Blazquez-Salcedo:2020rhf} for scalarized black holes with coupling function $f_2(\varphi_0)$ and $\beta=6$.}
\label{fig:loss_vs_error}}
\end{center}
\end{figure}

It is interesting, however, to notice that the final loss achieved by the optimizer in the last two cases is significantly larger than in the more accurate modes. We can therefore use this loss, which is nothing else than a measure of the degree to which the perturbation equation is satisfied, as a proxy for the success of the fit. This relationship between the final loss and the accuracy of the QNM frequencies is depicted in a log-log scale in FIG. \ref{fig:loss_vs_error}. Based on this plot, a final loss below 0.1 seems to indicate a reasonably accurate frequency. 

%%%%%%%%%%%%%%%%%%%%%%%%%%%%%%%%%%%%%%%%%%%%%%%%%%%%%%%%%%%%%%%%%%%%%%%%%%%%%%
\section{Results}
\label{sec:Results}
%%%%%%%%%%%%%%%%%%%%%%%%%%%%%%%%%%%%%%%%%%%%%%%%%%%%%%%%%%%%%%%%%%%%%%%%%%%%%%

\subsection{Quadratic Model}

\begin{figure}[thpb]
\begin{center}
\includegraphics[width=0.49\textwidth]{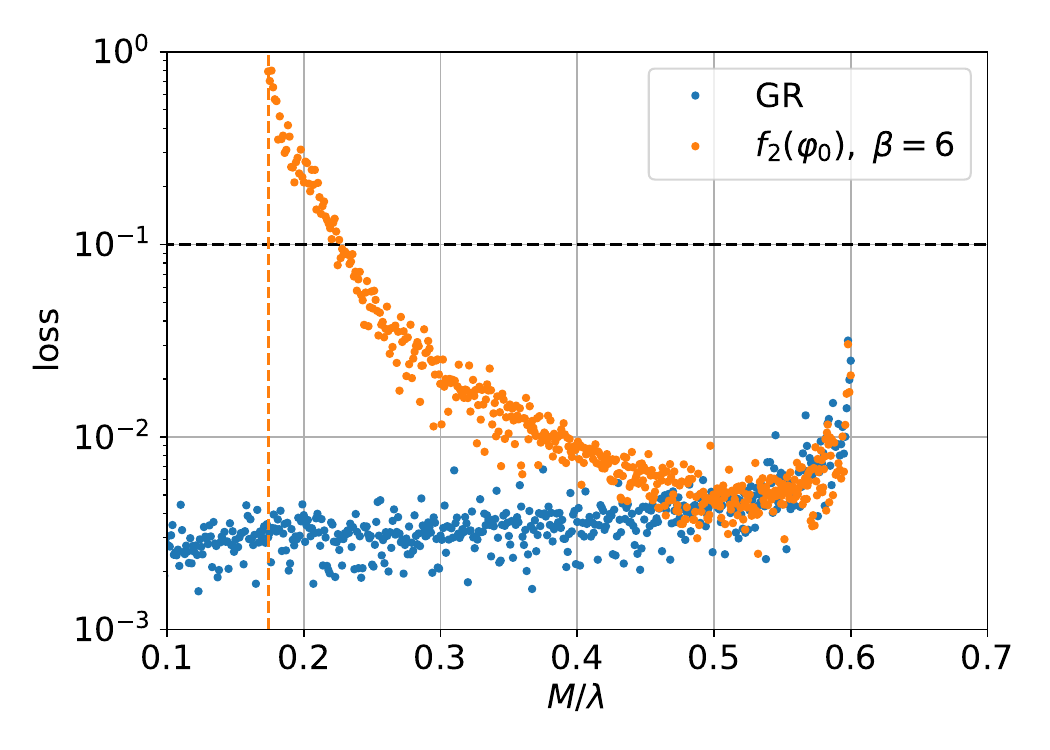}
\caption{Final training loss as a function of mass for GR and $f_2(\varphi_0)$ coupling function solutions. The vertical dashed line marks the point of hyperbolicity loss. The horizontal dashed line marks the threshold for the modes to be reliable.
\label{fig:Losses_CC18}}
\end{center}
\end{figure}

The first theory we analyze corresponds to the case of a quadratic coupling function $f_2(\varphi_0)$ given by Eq.~\eqref{eq:f_2} with $\beta = 6$. We focus on the fundamental axial QNM for $\ell=2$. In addition to the modes already compared in Table~\ref{tab:real_imag}, the much larger collection of solutions in this section can be qualitatively compared to the plots in \cite{Blazquez-Salcedo:2020rhf} as an additional check for the reliability of the PINNs as a QNM solver. 

\begin{figure}[thpb]
\begin{center}
\includegraphics[width=0.49\textwidth]{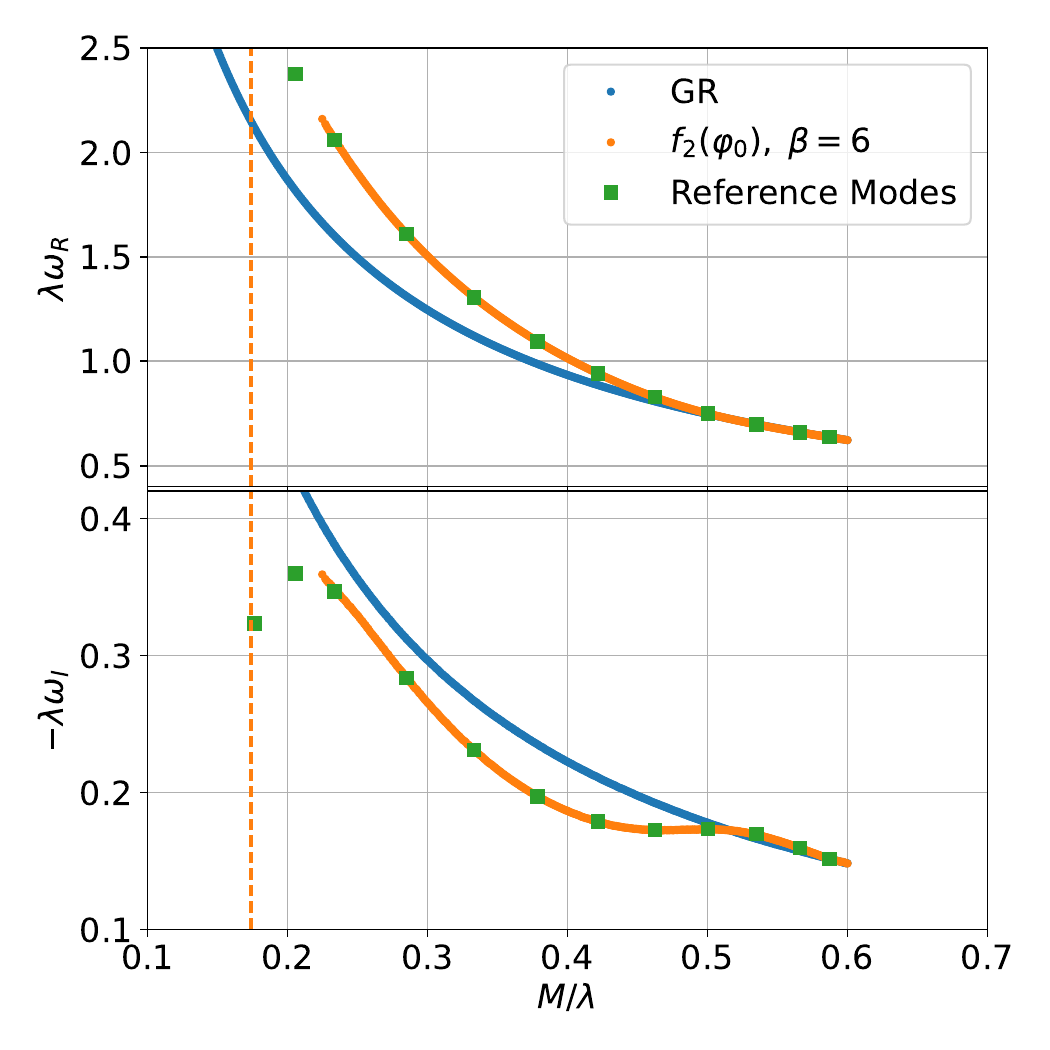}
\caption{Real (top) and imaginary (bottom) parts of the quasinormal modes for GR and the $f_2(\varphi_0)$ coupling function, as a function of $M/\lambda$. The green square dots correspond to the data from \cite{Blazquez-Salcedo:2020rhf}. The vertical dashed line marks the point of hyperbolicity loss. 
\label{fig:Modes_CC18_V1}}
\end{center}
\end{figure}

\begin{figure}[thpb]
\begin{center}
\includegraphics[width=0.49\textwidth]{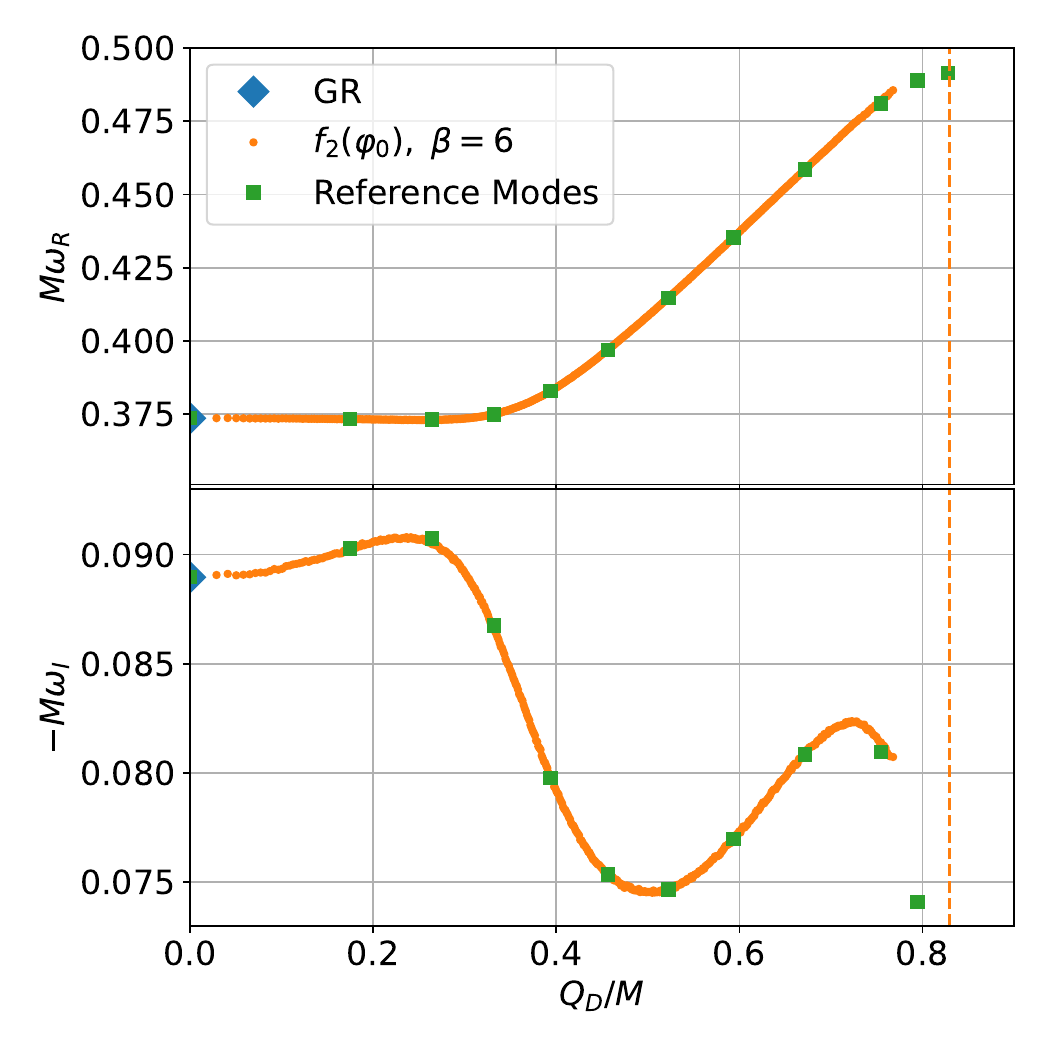}
\caption{Mass-normalized real (top) and imaginary (bottom) parts of the quasinormal modes for $f_2(\varphi_0)$ coupling function, as a function of $Q_D/M$. Only solutions with a final loss below 0.1 are plotted. The green square dots correspond to the data from \cite{Blazquez-Salcedo:2020rhf}. The GR limit is marked by a blue diamond. The vertical dashed line marks the point of hyperbolicity loss. 
\label{fig:Modes_CC18_V2}}
\end{center}
\end{figure}

The PINN has been trained on a family of 503 black hole solutions, with values of the horizon radius $r_H$ ranging from 0.196 to 1.2, in order to produce values for the quasinormal ringing frequencies for each of them. The quality of the QNM values is estimated by the value of the loss after the training of the PINN on each of the modes, as shown in FIG. \ref{fig:Losses_CC18}. As explained in Section \ref{subsec:testingPINN}, we have chosen a value of 0.1 as the maximum loss that is admissible for a reliable mode value. For this reason, only the modes with a final training loss below 0.1 will be represented in FIG. \ref{fig:Modes_CC18_V1} and \ref{fig:Modes_CC18_V2}. Comparison with the results of \cite{Blazquez-Salcedo:2020rhf} confirms that the threshold 0.1 for the loss is adequate.

FIG. \ref{fig:Modes_CC18_V1} shows the real and imaginary parts of the frequencies as a function of the black hole mass $M$. The frequencies are compared to the Schwarzschild modes, also computed by the PINN. When compared to the results of \cite{Blazquez-Salcedo:2020rhf}, the frequencies are remarkably accurate until approximately $M \approx 0.22$ (which corresponds to $r_H \approx 0.27$). Some specific values are listed in Table \ref{tab:real_imag}. Beyond this point, the QNM extracted by the PINN are no longer included in the plot as they start increasing rapidly and cannot be trusted. This is also confirmed by the PINN loss rapidly rising above the threshold of 0.1. As already mentioned, this could be caused by two reasons: On the one hand, by their proximity to the point $r_{S2}$, where the equations of motion are no longer hyperbolic. On the other hand, the ratio of the horizon size to the background solution resolution also becomes smaller.

FIG. \ref{fig:Modes_CC18_V2} relates the frequencies (multiplied by $M$) to the monopolar coefficient $Q_D$ (also known as the scalar charge) of the scalar field profile. These plots are particularly useful to distinguish small inaccuracies on the computation, as the simple inverse dependence of $\omega$ with $M$ is normalized away. Again, we observe that the modes agree well when compared to \cite{Blazquez-Salcedo:2020rhf} until the very close proximity of the hyperbolicity loss point.

\subsection{Quartic Model}

In this case, we take the quartic coupling function $f_4(\varphi_0)$ given by 
Eq.~\eqref{eq:f_4}. Again, we focus on the fundamental axial QNM for $\ell=2$. In this case, the PINN has been trained on families of background solutions with $\beta = 400, 1000 \text{ and } 10000$ respectively, as summarized in TABLE \ref{tab:f4_cases}. 
\begin{table}[H]
\begin{center}
  \begin{tabular}{|c||c|c|c|}
    \hline
    $\beta$ & \# of cases & min $r_H$ & max $r_H$ \\ \hline \hline
    400     & 319         & 0.158     & 0.794     \\ \hline
    1000    & 275         & 0.082     & 0.630     \\ \hline
    10000   & 180         & 0.016     & 0.352     \\ \hline
    \multicolumn{4}{c}{} 
  \end{tabular}
  \caption{Summary of black hole background solutions considered for the $f_4(\varphi_0)$ models. \label{tab:f4_cases}}
\end{center}
\end{table}
\begin{figure}[h]
\begin{center}
\includegraphics[width=0.49\textwidth]{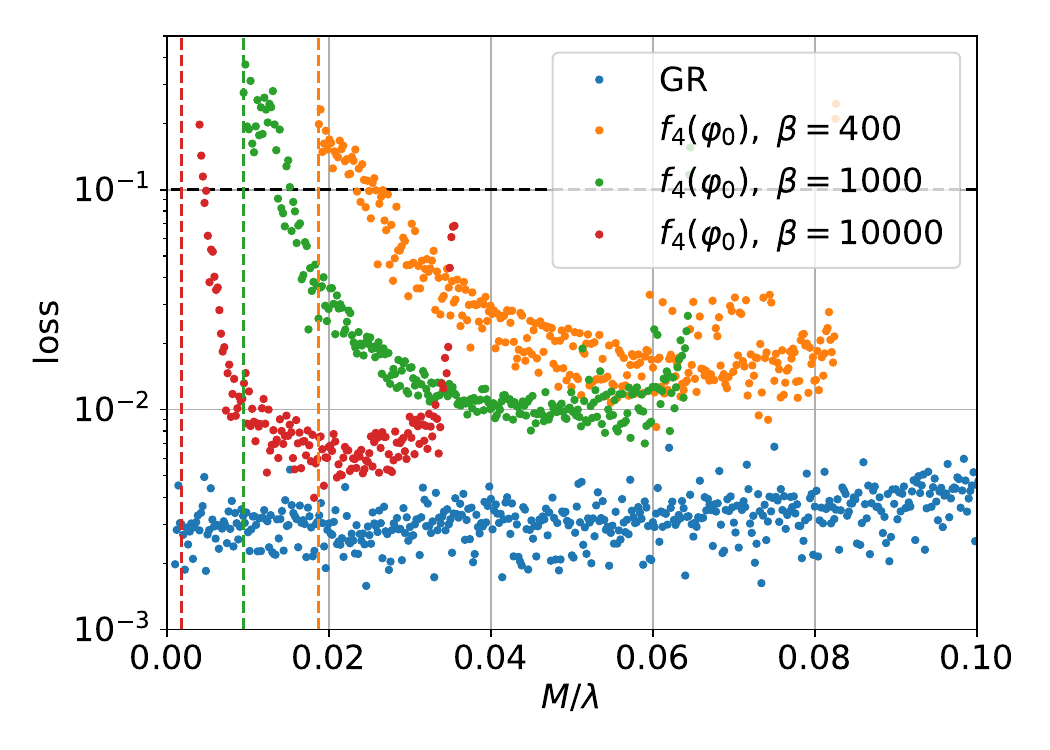}
\caption{Final loss for GR and $f_4(\varphi_0)$ coupling functions. The vertical dashed lines mark the points of hyperbolicity losses. The horizontal dashed line marks the threshold for the modes to be reliable.
\label{fig:Losses_CC16}}
\end{center}
\end{figure}
\begin{figure}[thpb]
\begin{center}
\includegraphics[width=0.49\textwidth]{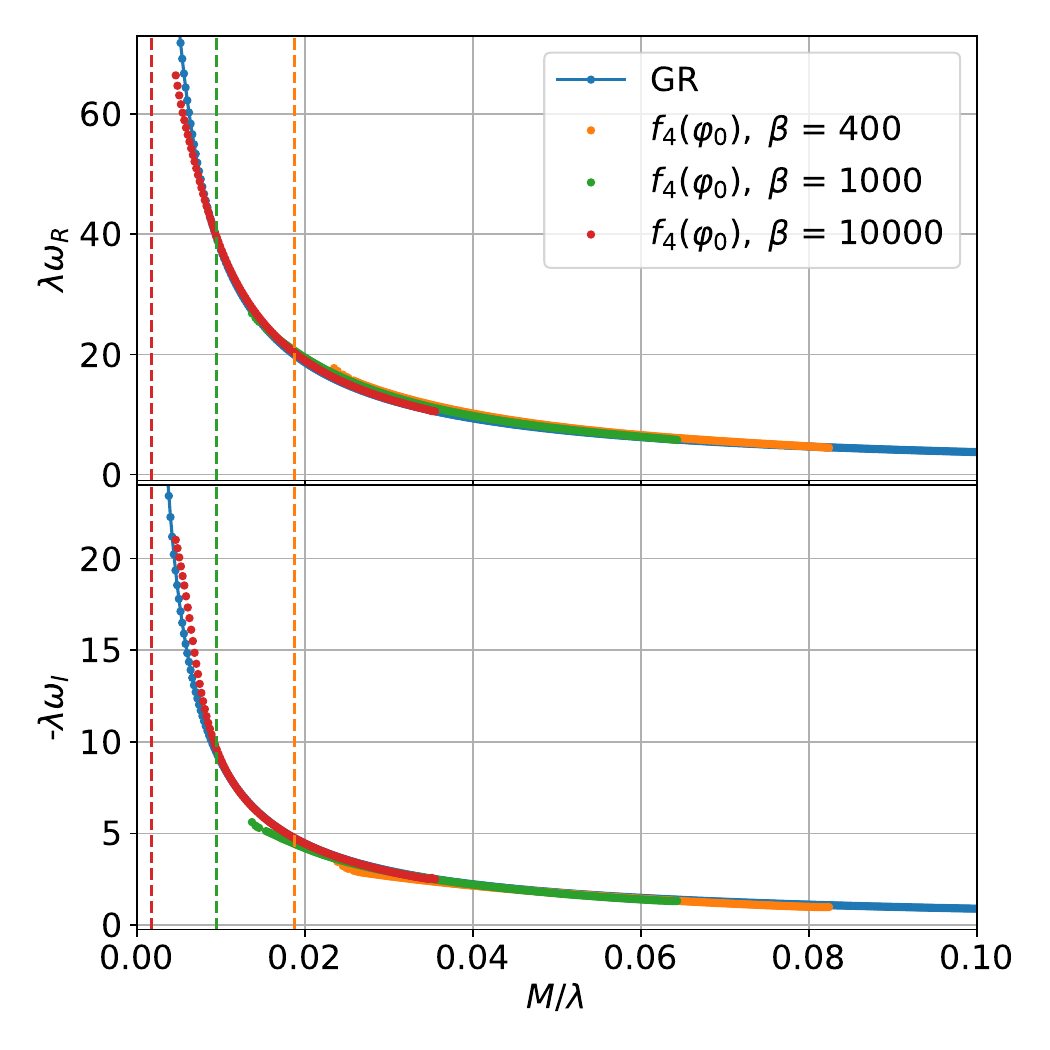}
\caption{Real (top) and imaginary (bottom) parts of the quasinormal modes for GR and the $f_4(\varphi_0)$ coupling function, as a function of $M/\lambda$. The vertical dashed lines mark the points of hyperbolicity loss of each theory.
\label{fig:Modes_CC16_V1}}
\end{center}
\end{figure}

The particular choices of $\beta$ are made such that we have a large enough hyperbolic part of the branch \cite{Blazquez-Salcedo:2022omw}\footnote{{Typically for nonlinearly scalarized black holes, with the decrease of $\beta$ the deviation from GR increases but also the black hole solutions become non-hyperbolic against linear perturbation \cite{Blazquez-Salcedo:2022omw}.}}.  

\begin{figure}[thpb]
\begin{center}
\includegraphics[width=0.49\textwidth]{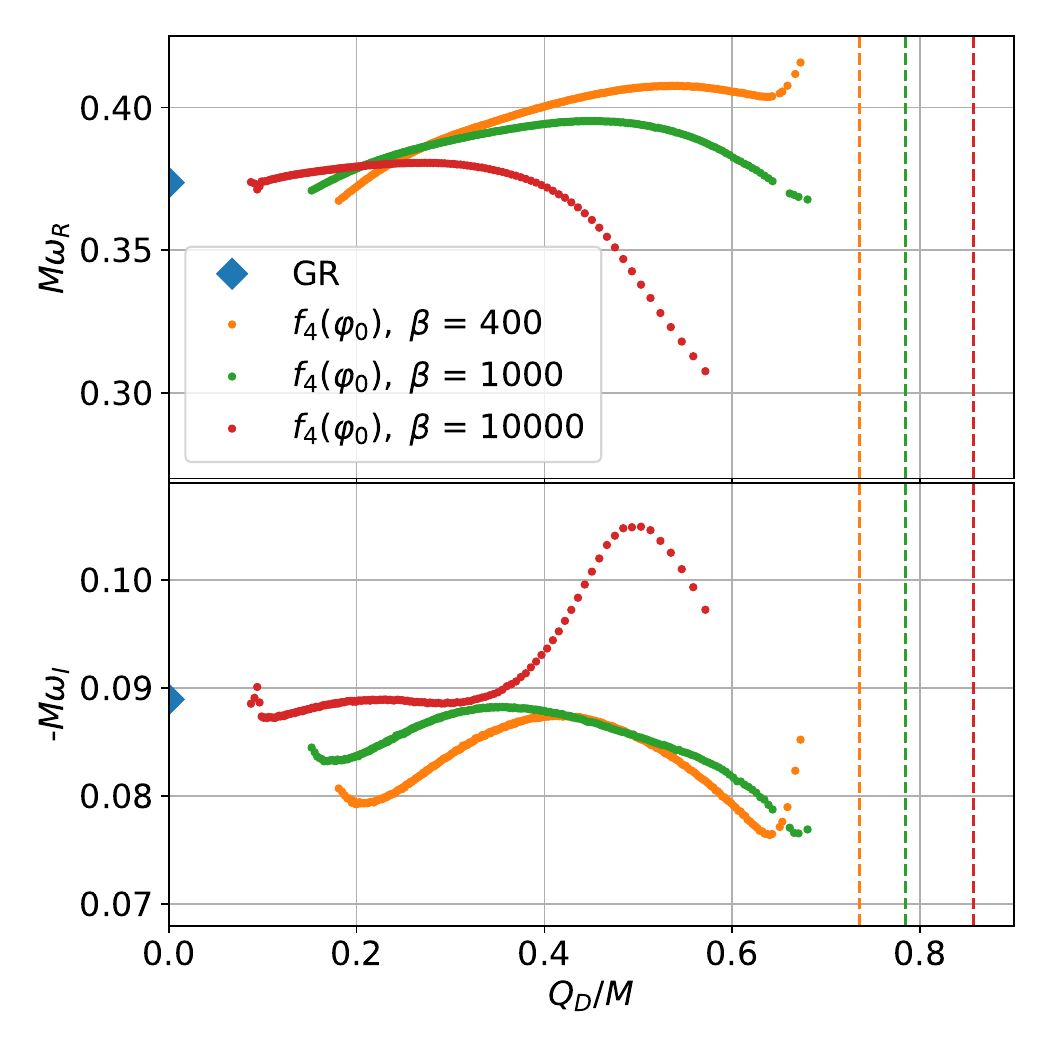}
\caption{Mass-normalized real (top) and imaginary (bottom) parts of the quasinormal modes for the $f_4(\varphi_0)$ coupling function, as a function of $Q_D/M$. The GR limit is marked by a blue diamond. The vertical dashed lines mark the points of hyperbolicity loss of each theory, and the modes closest to them may be less accurate.
\label{fig:Modes_CC16_V2}}
\end{center}
\end{figure}

FIG. \ref{fig:Losses_CC16} shows the final losses after the training of the PINN in each of the modes, also including GR as a reference. Again, we take 0.1 as the threshold for the solutions to be trustworthy, and as such only the ones below the threshold are represented in FIG.~\ref{fig:Modes_CC16_V1} and \ref{fig:Modes_CC16_V2}.

We plot the frequencies of the three families of solutions as a function of the total mass $M$ in FIG. \ref{fig:Modes_CC16_V1}, together with the Schwarzschild solutions as a reference. In this case, all models are hard to distinguish from each other, as the frequencies are very similar and the main variation is dominated by the inverse relation to $M$. The differences between models are much more clear when plotting the combination $\omega M$, as shown in FIG. \ref{fig:Modes_CC16_V2}, as a function of the mass-normalized scalar charge $Q_D/M$. 

A natural question to ask is whether the obtained deviations from GR  are actually measurable. The most promising source in this context is the post-merger ringing that will be observable with great accuracy (reaching signal-to-noise ratio over 100) by the next generation of gravitational-wave detectors Einstein Telescope, Cosmic Explorer, and LISA. Results based on the parameterized QNM test demonstrate that a change in the QNM frequency with respect to GR can be constrained as low as $0.1\%$ for a non-rotating black hole \cite{Maselli:2019mjd}. However, the post-merger remnant black hole is theorized to be rapidly rotating, which would eventually worsen these estimates by roughly an order of magnitude. Nevertheless, as one can see in Fig. \ref{fig:Modes_CC16_V2}, for small black hole masses the difference between GR and scalarized black hole QNMs  reaches up to $\sim 18\%$ in the case with $\beta = 10^4$. Even though this is certainly measurable according to the estimates in \cite{Maselli:2019mjd} for an idealized setup, the post-merger remnant offers a number of complications, such as rapid rotation, contamination by the nonlinear part of the merger, etc., that worsen the picture. Therefore, further studies are needed to analyze up to what extent the deviations reported in this work give a measurable deviation in the observed ringdown GWs.

%%%%%%%%%%%%%%%%%%%%%%%%%%%%%%%%%%%%%%%%%%%%%%%%%%%%%%%%%%%%%%%%%%%%%%%%%%%%%%
\section{Discussion}
\label{sec:Discussion}
%%%%%%%%%%%%%%%%%%%%%%%%%%%%%%%%%%%%%%%%%%%%%%%%%%%%%%%%%%%%%%%%%%%%%%%%%%%%%%

In the present paper we have explored the power of physics informed neural networks (PINNs) to calculate quasinormal modes of black holes beyond general relativity. Previous similar studies in GR~\cite{Luna:2022rql} have been here extended in two main directions. First, the complexity of the perturbation equations one has to solve is increased significantly, and new features appear such as hyperbolicity loss for small mass black holes. Second, the background solution is known only numerically. All this requires an increase in the complexity of the employed neural network and makes the convergence to the true QNM frequencies more subtle. 
	
The modified theory of gravity we have focused on is Einstein-scalar-Gauss-Bonnet (EsGB) gravity. This theory is well-motivated and offers rich black hole phenomenology beyond GR such as violation of the no-scalar hair theorems and black hole scalarization. Black hole perturbations have already been explored in the literature for different subclasses of this theory, helping us build intuition. Therefore, EsGB gravity is an ideal playground for applying the PINN approach and testing its power/accuracy. 

As a good more controllable starting point we have explored the axial perturbations of scalarized black holes within this theory where the perturbation equations for the metric and the scalar field reduce to a single master equation. The coupling between the scalar field and the Gauss-Bonnet invariant has been chosen to have either a quadratic or a quartic form. The former case corresponds to the standard black hole scalarization where the Schwarzschild  black hole becomes unstable for large scalar field coupling strength giving rise to a linearly stable hairy black hole solution. The QNMs of such black holes have already been explored in the literature \cite{Blazquez-Salcedo:2020rhf} and they were used for calibration of the developed PINN. Correspondingly, for the quartic coupling the Schwarzschild black hole is always linearly stable but can co-exist with hairy black holes which are also linearly stable and thermodynamically more favorable. The axial QNMs have not been examined before in the literature for this case.

The PINN that we have used in this work to compute the QNM frequencies is a very simple feedforward fully connected neural network that is able to approximate the eigenfunctions of the perturbation equation and extract the QNM. The machine learning environment \texttt{PyTorch} makes the implementation of the network remarkably simple as it provides automatic differentiation capabilities, and allows the training to be easily accelerated in one or multiple GPUs. The weights, biases and frequency of a given mode can also be stored and used as a starting point for the training on neighboring background solutions, which makes the convergence even faster. Thanks to all these features, each mode can be accurately computed in about 15 seconds on a standard GPU, with the possibility of training different families of solutions simultaneously.

We have validated the accuracy of the developed PINN against the already available results in \cite{Blazquez-Salcedo:2020rhf}, where we have checked that the error in the calculated QNM frequencies is less than 0.1\% for most of the parameter range. This also includes black hole background models for which the effective potential of the perturbation equation has negative parts. Therefore, PINNs are powerful tools even for more complicated potential forms.
	
The error of the calculated QNM frequencies with PINNs is somewhat more difficult to control compared to conventional numerical methods, as the computed results do not have a clearly defined convergence rate as the resolution is increased. Even more so, the relation between the network training loss and the QNM accuracy does not have a simple theoretical form. Nevertheless, after performing some experiments and comparison with known results, we have determined a rough criteria for an accuracy metric, namely that the loss at the end of the training should be smaller than 0.1. This condition breaks only as the black hole sequences approach the point where the hyperbolicity of the PDE is lost, i.e.~the coefficient in front of the second order time derivative changes sign. 
	
After validating our methodology, we have presented results for a quartic coupling between the scalar field and the Gauss-Bonnet invariant. The black hole solutions in this case are characterized by a very interesting property -- both the Schwarzschild black hole and the scalarized ones are linearly stable. In addition, the two branches of solutions can only meet at the zero mass point. Therefore, unlike the scalarized black holes with quadratic coupling above, there is no bifurcation point where the GR solutions merge with the scalarized ones. The calculated QNM frequencies in this case also show a smooth behavior with a final training loss consistently being under $0.1$ if we are sufficiently far away from the point of hyperbolicity loss.
	
Overall, our results demonstrate the power of PINNs to calculate accurate QNMs for  more complicated cases than pure Schwarzschild or even Kerr black holes. It will be interesting to further investigate if PINNs can perform equally well for even more involved systems of differential equations in beyond-GR theories. A natural extension of the work presented here is to calculate the polar QNMs of scalarized black holes where one ends up with a system of ordinary differential equations. Note that for black holes in EsGB gravity, the degeneracy between polar and axial QNMs is broken and instead, we have two families of distinct QNM frequencies. Those results will be reported elsewhere.

%%%%%%%%%%%%%%%%%%%%%%%%%%%%%%%%%%%%%%%%%%%%%%%%%%%%%%%%%%%%%%%%%%%%%%%%%%%%%%
\begin{acknowledgments}
%%%%%%%%%%%%%%%%%%%%%%%%%%%%%%%%%%%%%%%%%%%%%%%%%%%%%%%%%%%%%%%%%%%%%%%%%%%%%%

We thank Alejandro Torres-Forn\'e for very useful discussions, and Jos\'e Bl\'azquez-Salcedo for providing the data used in Table \ref{tab:real_imag}.
RL acknowledges financial support provided by Generalitat Valenciana / Fons Social Europeu through APOSTD 2022 post-doctoral grant CIAPOS/2021/150.
DD and SY acknowledge the  support by the Bulgarian NSF Grant KP-06-H62/6.
DD acknowledges financial support via an Emmy Noether Research Group funded by the German Research Foundation (DFG) under grant no.~DO 1771/1-1.
JAF is supported by the Spanish Agencia Estatal de Investigaci\'on (grant PID2021-125485NB-C21) funded by MCIN/AEI/10.13039/501100011033 and ERDF A way of making Europe, and the Generalitat Valenciana (grant CIPROM/2022/49). We also acknowledge further support by the European Horizon Europe staff exchange (SE) programme HORIZON-MSCA2021-SE-01 Grant No. NewFunFiCO-101086251.
This work has used the following open-source packages: \texttt{PyTorch}~\cite{PyTorch}, \texttt{Pandas}~\cite{Pandas}, \texttt{NumPy}~\cite{harris:2020}, \texttt{SciPy}~\cite{scipy:2020} and \texttt{Matplotlib}~\cite{Hunter:2007}.

%%%%%%%%%%%%%%%%%%%%%%%%%%%%%%%%%%%%%%%%%%%%%%%%%%%%%%%%%%%%%%%%%%%%%%%%%%%%%%
\end{acknowledgments}
%%%%%%%%%%%%%%%%%%%%%%%%%%%%%%%%%%%%%%%%%%%%%%%%%%%%%%%%%%%%%%%%%%%%%%%%%%%%%%

\bibliography{references}

\end{document}